# Superconducting detector magnets for high energy physics

**Matthias Mentink**[a,1], **Ken-ichi Sasaki**[b,2], **Benoit Cure**[a], **Nikkie Deelen**[a], **Alexey Dudarev**[a], **Mitsushi Abe**[b], **Masami Iio**[b], **Yasuhiro Makida**[b], **Takahiro Okamura**[b], **Toru Ogitsu**[b], **Naoyuki Sumi**[b], **Akira Yamamoto**[a,b], **Makoto Yoshida**[b] and **Hiromi Iinuma**[c]

[a] *CERN,*
  *1211 Geneva 23, Switzerland*

[b] *KEK,*
  *1-1 Oho, Tsukuba, Japan*

[c] *Ibaraki University,*
  *162-1 Shirakata, Tokai-mura, Japan*

[1] *E-mail*: `matthias.mentink@cern.ch`

[2] *E-mail*: `ken-ichi.sasaki@kek.jp`

## Executive Summary

Various superconducting detector solenoids for particle physics have been developed by every institute in the world since 1970's. The key technology is the aluminum-stabilized superconducting conductor for almost all the detector magnets in particle physics experiments. With the progress of the conductor, the coil fabrication technology has progressed as well. The outer support cylinder is used to support large hoop stress, and the conductor is directly wound inside the cylinder. Indirect cooling is usually adopted in order to reduce materials used in magnet structure and maximize the transparency for charged particles passing through by eliminating the liquid helium vessel. Pure aluminum strips are frequently used as temperature equalizer and fast quench propagator to give a uniform temperature rise during quench. The vacuum vessel design study has also progressed, especially for transparent detector solenoids through isogrid honeycomb technology for outer vacuum vessel wall. These technologies have been used successfully for the manufacturing of various superconducting detector magnets in the past four decades thanks to the technical and scientific competencies developed with various breakthroughs.

The detector solenoids design study is in progress for future big projects in Japan and Europe, that is, ILC, FCC and CLIC, based on the technologies established over many years. The magnet size for each project is as large as or larger than the magnets developed for CMS and ATLAS-CS, and the combination of good mechanical properties and keeping a high RRR is a key point for the development of Al-stabilized conductor. In addition, a larger current capacity is required to accommodate the larger bore size and the magnetic volume. The ILC and FCC groups are continuing the design study of the conductor.

The present concern for the detector solenoid development is to have been gradually losing the key technologies and experiences, because large-scale detector magnets with Al-stabilized conductor has not been fabricated after the success of CMS and ATLAS-CS in LHC in 1990s – 2000s. Complementary efforts are needed to resume an equivalent level of expertise, to extend the effort on research and to develop these technologies and apply them to future detector magnet projects. Especially, further effort is necessary for the industrial technology of Al-stabilized superconductor production. The worldwide collaboration with relevant institutes and industries will be critically important to re-realize and validate the required performances. KEK and CERN jointly propose to organize a workshop to share the awareness of industrial issue on Al-stabilized conductor fabrication, and to invite superconducting magnet scientists, engineers of conductor industries and physicists who plan and design the future particle experiments.

Some detector solenoids for mid-scale experiment sometimes use a conventional Cu-stabilized Nb-Ti conductor. The conventional solenoids for such experiments do not need the development of unique conductor, but instead, precise control of magnetic field distribution would be required. The development efforts are on-going in terms of the magnetic field design technology with high precision simulation, the coil fabrication technology to achieve the design requirement and the control method of magnetic field distribution.

## 1. Introduction

A superconducting detector magnet is one of the key components for particle physics experiments to analyze the momentum and polarity of charged particles. It is required to have a large warm bore to install many types of particle detectors, and a large solid angle to maximize the detection efficiency of particles. Many magnets have been developed since 1977 [1]. Table 1 summarizes the advances in thin detector solenoids. FigureFig. 1.1 shows general parameters and configuration of the ATLAS-CS and CMS detector solenoids at the CERN LHC experiments representing most recent advances [1].

Design studies of superconducting detector solenoids are progressing for the several future projects. The target central field is generally 2 – 5 T, which is mainly determined by the resolution of particle detector systems. Superconducting detector magnets can be roughly categorized into two types, considering the required feature based on the arrangement of calorimeter position. One is a transparent magnet and another is a non-transparent magnet. The transparent magnet, like ATLAS-CS etc., requires high transparency for charged particles passing through. Calorimeters are placed outside the detector magnet and therefore the charged particle needs to pass through the coils and its cryostat with a minimum energy loss as small as possible. On the other hand, in non-transparent magnets, like CMS etc., all the particle detectors are generally installed inside the magnet bore except for muon detectors. This results in much larger bore and much longer coil length than those of the transparent magnet. An important feature is a generally higher magnetic field, with less constraint with the transparency, resulting in a large operational current.

The common developing item for both type magnets is the conductor combining both high-strength and low-resistivity Al stabilizer and conductor development efforts are ongoing in the worldwide projects, CLIC, ILC and FCC (-ee, -hh). Other development efforts, like the coil winding, quench protection and over all structural design technologies are also being progressed according to the requirements of detector solenoids. The present status and future prospects of development items are summarized, and the present design status of the magnets for future projects are reported in the latter section.

Table 1.1 Advances in thin/transparent solenoid magnet technology.

| Technology | First Detectors of the technology implemented |
|---|---|
| Al-stabilized superconductor (soldered) and indirect/conduction cooling | ISR[2], CELLO[3] |
| Secondary winding and quench back | PEP4-TPC[4] |
| Co-extruded Al-stab. superconductor | CDF[5] |
| Inner winding | TOPAZ[6] |
| CFGP outer vacuum vessel/wall | VENUS[7] |
| Thermo-siphon and indirect cooling | ALEPH[8], DELPHI[9] |
| 2-layer coil and grading | ZEUS[10], CLEO[11] |
| Al-stabilizer w/ Zn, and Isogrid vacuum vessel | SDC-Prototype[12] |
| Shunted coil w/ conductor soldered to mandrel | CMD-2[13] |
| High-strength Al-stabilizer w/ Ni micro-alloying and fast quench propagation w/ pure-AL strips and heater | ATLAS[14] |
| Hybrid conductor configuration using EBW | CMS[15] |
| Self-supporting coil with no outer support cylinder | BESS-Polar[16] |

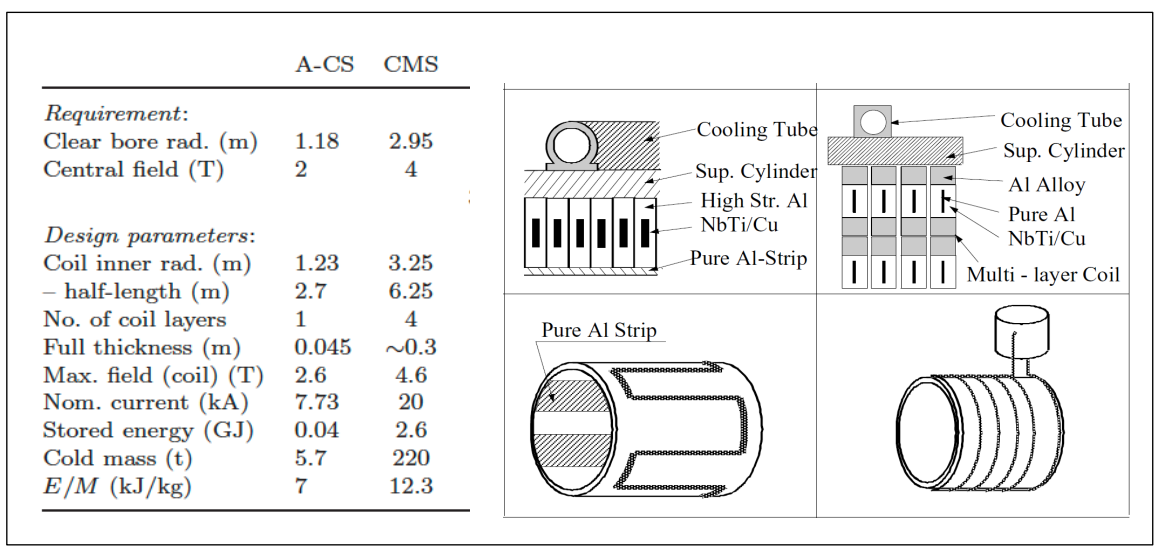

Fig. 1.1 Detector solenoids experienced in LHC

## 2. Technology for detector magnets

Many engineering efforts have been made to improve transparency of the detector solenoids, and modern detector solenoid design concepts have been realized [17].

- A superconducting coil is wound with Nb-Ti/Cu conductor/cable cladded with pure aluminum stabilizer.
- The coil is cooled with its own thermal conductivity from cooling pipes set on an outer support cylinder.
- No structure (i.e. bobbin) exists inside the coil winding.
- Epoxy based resin is painted or impregnated into the coil winding to integrate winding conductors and the outer support cylinder both mechanically and thermally.

These design concepts give good transparency for the particles passing through, as well as a light cold-mass weight and a simple coil structure. Consequently, coils for non-transparent solenoids have been designed with similar design approaches, too. An overview of the recent developments for the detector solenoids is described below.

## 2.1 Aluminum stabilized superconductor and superconducting coil

Aluminum stabilization of the superconductor is a key technology in modern detector magnets. It contributes to the stability of the superconductor with minimum material and weight.

The electro-magnetic forces generated in the coil winding are sustained by the conductors themselves in combination with the outer support cylinder. Since pure aluminum stiffness is rather low with a yield strength of about 30 MPa, the outer support cylinder made of aluminum alloy would need to contribute enough mechanical strength to keep the stress in the coil at a reasonable level. This means that reinforcement of conductor saves thickness of the outer support cylinder. Aluminum-stabilized superconducting conductors have benefitted from a number of improvements, notably regarding its mechanical strength over the last four decades. The evolution of conductors is summarized in Fig. 2.1. One approach has been to provide homogeneous reinforcement of the stabilizer itself; the other was to work with a hybrid configuration of soft high conductivity material with a strong alloy.

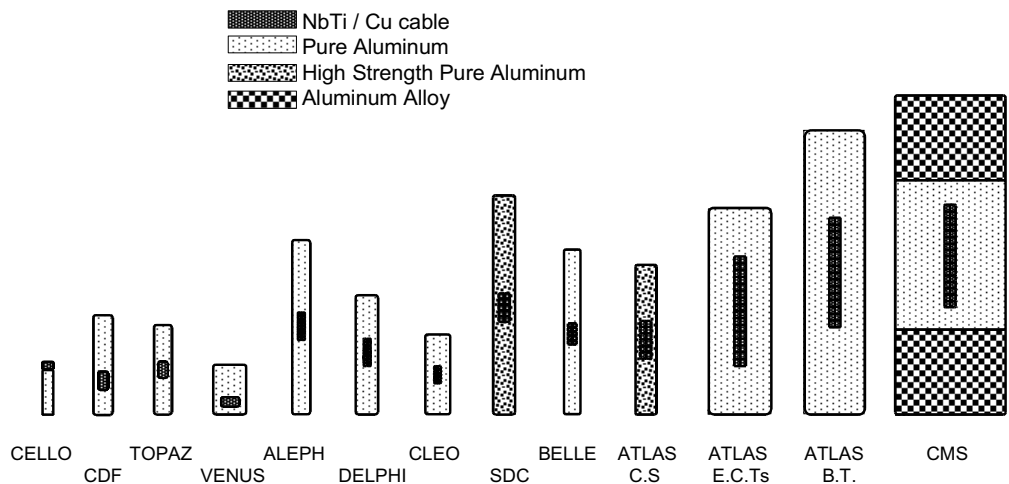

**Fig. 2.1** Evolution of conductors for detector magnets. From left to right: CELLO, CDF, ALEPH, ATLAS-CS, -BT and CMS.

Homogeneous reinforcement was established by combining micro-alloying and cold-work hardening. It was found that nickel additive effectively contributes to mechanical strength while keeping a reasonably low electrical resistivity in the aluminum. FigureFig. 2.2 shows progress of mechanical strength of aluminum stabilizer as a function of electrical resistance at 4.2 K in comparison with typical oxygen free copper (OFC) as stabilizer. It can be seen that the strength of the aluminum stabilizer has become comparable with that of copper, while maintaining its high RRR and the all-important advantage of lightness. The conductor clad with reinforced 0.5 % Ni aluminum stabilizer gives a one-third reduction in the thickness of the ATLAS central solenoid to when compared to a conductor using a pure aluminum stabilizer [18][19].

A hybrid configuration, which consists of a combination of pure aluminum stabilized superconductor with high strength aluminum alloy (A6082) blocks attached to both sides by electron-beam welding was developed for the CMS solenoid. Such a hybrid configuration is very effective in large-scale conductors because it can be welded. It allows a hoop strain of 0.15 % induced by a hoop stress of 105 MPa, and it is an essential feature of the 4 T CMS solenoid design.

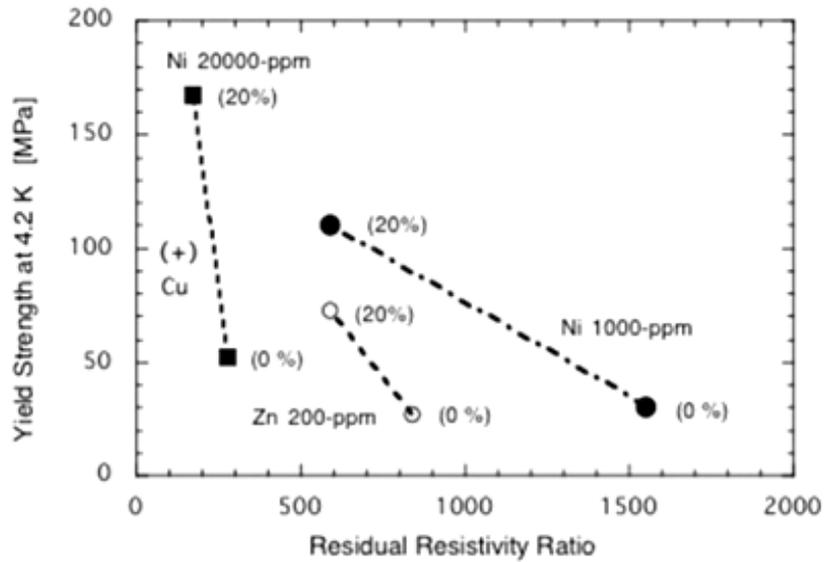

**Fig. 2.2** The progress made in the development of high-strength aluminum having good electrical conductivity.

In the hybrid approach the electro-magnetic force acts on the superconductor, which is confined in the soft pure aluminum. In order to ensure that the conductor does not migrate in this medium when it is required to operate at fields greater than 4 T that are being considered for future detectors, it is envisaged to combine the two approaches to reinforcement, co-extruding the conductor with the micro-alloyed material followed by electron beam welding (EBW) of the tough alloy flanges. Table 2.1 summarizes the relevant parameters of high-strength conductors [19].

**Table 2.1** Relevant parameters of high-strength conductors.

| Type | Composition | Yield strength (MPa) | | RRR |
|---|---|---|---|---|
| | | Al | Full conductor | |
| ATLAS-CS | Ni(0.5%)Al | 110 | 146 | 590 |
| CMS | Pure Al & A6082-T6 | 26 428 | 258 | 1400 |

## 2.2 Inner winding technique to support cylinder and indirect/conduction cooling

A traditional way of the solenoid coil winding is to wind the conductor with tension to the outside of a mandrel/bobbin. The tension should be sufficiently large to ensure compressive pre-stress between the coil-winding and inner-bobbin for eliminate the separation of the coil from the inner-bobbin when the hoop stress increases in the conductors according to the coil excitation. This requires the bobbin to be thick enough to avoid buckling. Conversely, if the winding can be done inside a support cylinder then the compressive force further increases when the current is increased. There is no bucking force during the winding. In addition, having the ground insulation between coil and the support cylinder under pressure ensures the good thermal conduction required for indirect cooling from the cooling pipe placed on the outer surface of the support cylinder. LHe flow may be realized by using the force 2 phase helium flow in the cooling pipe attached to the outer support cylinder, or by natural gravitational convention inside the cooling pipe configured for enabling it [1][9][10][13][14]. As an example, ATLAS-CS winding concept is shown in Fig. 2.3.

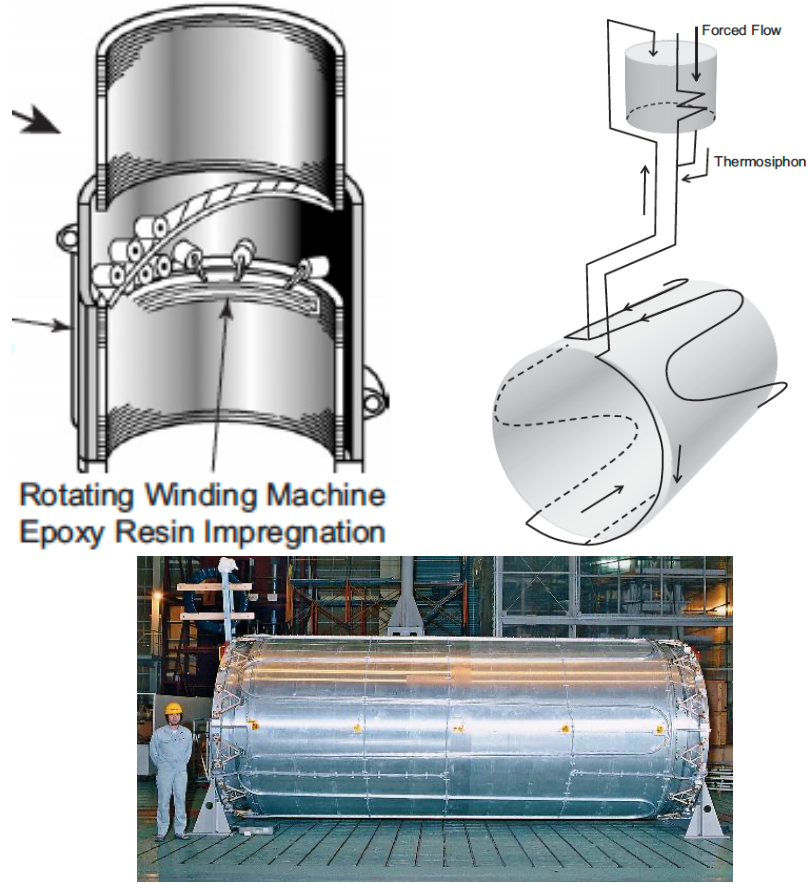

**Fig. 2.3** Concepts of inner coil winding and indirect/conduction cooling of the detector solenoid with a photo from the ATLAS-CS support cylinder with cooling pipe.

## 2.3 E/M ratio and transparency

Compactness and transparency of the magnet are important in order to create a magnetic field with minimum disturbance for the particles and having maximum detector acceptance. For these reasons, the ratio of stored energy to effective coil cold mass, called the *E/M* ratio, is a useful parameter to scale the lightness, and compactness (or efficiency) of the magnet.

In the case of a solenoid coil,

$$E/M = \left.\int \frac{B^2}{2\mu_0} dv \right/ \rho V_{coil} = \sigma_h/2\rho$$

Where $\sigma_h$ is hoop stress, $\rho$ equivalent coil density. The necessary coil thickness is determined by

$$t = \frac{B^2}{2\mu_0}\frac{1}{\sigma_h}$$

The E/M ratios in various detector solenoid magnets are shown in Fig. 2.4 [1][19]. In early generations of thin magnets a typical E/M ratio was 5 kJ/kg. Based on the development of high strength aluminum stabilizer, ~10 kJ/kg was achieved for the SDC prototype. Using similar high-strength aluminum stabilizer, the ATLAS central solenoid reached 8.1 kJ/kg at its test field of 2.1 T. The CMS solenoid achieved an E/M ratio of 12 kJ/kg at its nominal field of 4.0 T. While it was not required to be a thin solenoid, there was a strong incentive to moderate the mass of the coil for reasons of physical size as well as cost containment. During testing, a prototype magnet for the BESS Polar program achieved E/M ≈13 kJ/kg without damage.

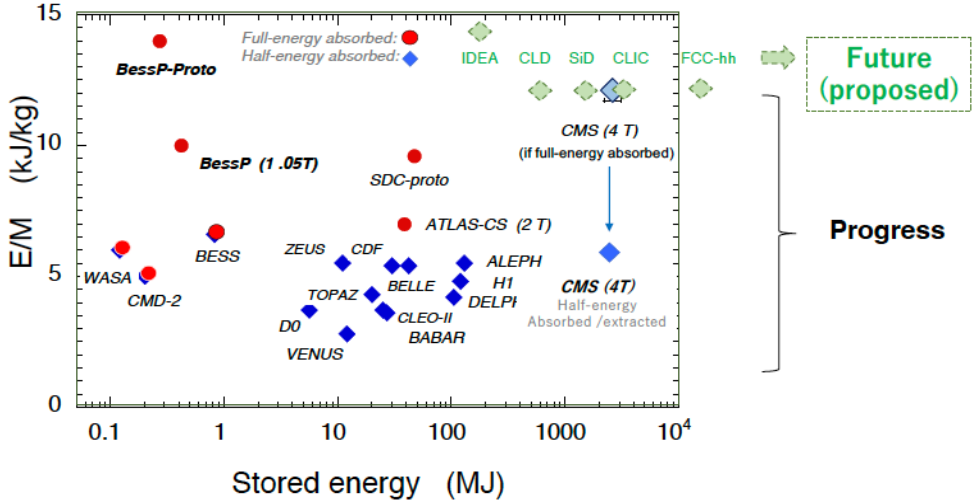

**Fig. 2.4** E/M ratio as a function of stored energy (E). Diamond (blue) plots show half energy to be absorbed in the coil with E/M level of ~ ≤ 5 kJ.kg, and round plots show full energy absorbed in the superconducting coil. Diamond (light green) plots show future solenoids proposed.

Another limiting factor for the E/M ratio is the quench protection. In the adiabatic condition temperature rise after quench can be expressed as,

$$\int_{t_{quench}}^{t_{end}} j^2 dt = \int_{T_0}^{T_{max}} C_{pave}/\rho_{ave} \, dT$$

where $t_{quench}$ is the time quench occur, $t_{end}$ is the time current is completely down, $j$ is the current density, $T_0$ is the operation temperature, $T_{max}$ is the maximum temperature, $C_{pave}$ is the average volumetric specific heat of the conductor, and the $\rho_{ave}$ is the average electric resistivity of the conductor. The equation indicates that once the conductor material is determined the maximum temperature after quench is proportional to the square of current density times the current dischareg time (i.e., the concept so called MIITs) [20]. Since the detector solenoid has large stored energy and also contains sensitive detector electronics in its aperture, the discharge time may not be shortened too much. For the large-scale conduction cooling solenoid, to avoid the excess thermal stress in the structure, it is generally required to limit the maximum temperature to relatively low value with a range of < ~150K. From these reasons engineering current density of the detector solenoid is generally kept low. To achieve the transparent solenoid magnet with such low current density, use of Aluminum stabilizer is essential with its light weight and good electric conductivity.

## 2.4 Thermal stabilization and fast quench propagation by using pure-Al strips

An effort to improve thermal stabilization and fast quench propagation by using pure-Al strips was proposed to suppress the maximum temperature expected with the MIIT concept described above. It will become further helpful to homogenize the coil temperature in case of the energy extraction system not working, the full stored energy needs to be absorbed with less peak temperature and as uniformly as possible in the coil [16][17][21]. This is possible if the quench propagation is much faster than the power decay time during a quench. A technique to increase the quench propagation velocity is to use axial pure-aluminum strips, as the concept shown in Fig. 2.5. This idea was implemented and experimentally verified in the development of various thin superconducting solenoids [14][16][17].

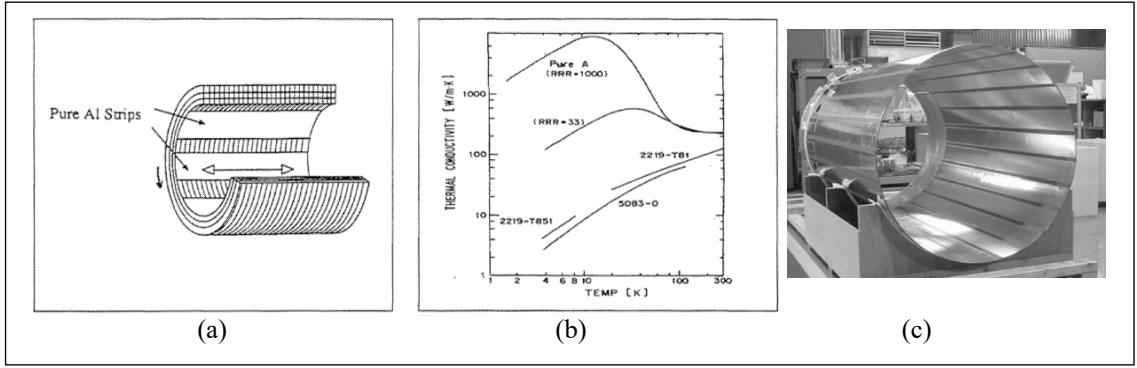

(a)　　　　　　　　　　(b)　　　　　　　　　　(c)

**Fig. 2.5** (a) Axial pure-Al strip quench propagator to enhance the effective thermal conductance along the coil axial direction, and (b) thermal conductivity of pure Al strip compared with other materials, (c) a photo of coil inner surface covered with pure Al strips along the coil axis for the BESS-Polar solenoid.

If an adiabatic condition was assumed, the longitudinal (along the conductor) quench propagation velocity $V_\phi$ is given by:

$$V_\phi = (J / \gamma C) \cdot \{L_0 T_s / (T_s - T_0)\}^{1/2}$$

where J is the current density, $\gamma C$ is the volumetric specific heat, $L_0$ is the Lorentz number, $T_s$ is the wave-front temperature and $T_0$ is the initial operational temperature. The relative axial (transverse) velocity is given by:

$$V_z = \{k_z / k_\phi\}^{1/2} \cdot V_\phi$$

where $V_z$ and $V_\phi$ are axial and circumferential quench velocities, respectively, and $k_z$ and $k_\phi$ are axial and circumferential thermal conductivity, respectively. One sees that $v_x$ may be enhanced by increasing the axial thermal conductance $k_z$. Normally, the axial thermal conductance is suppressed by turn-to-turn electrical insulation made of Kapton (Upilex) and/or glass tap. A pure aluminum strip of 1-2 mm thickness glued on inner surface of the coil serves to enhance the effective thermal conductance in the axial direction by bypassing the axial electrical insulation. As shown in Fig. 2.5, pure aluminum with RRR ≥ 1000 is especially appropriate for this purpose because of its enhanced thermal conductivity around the temperature range in us. At 4.5 K, $k_\phi$ is > 2000 W/m•K. With the 1-2 mm thick pure aluminum strip, $k_z$ is estimate to be ~ 100 W/m•K effectively with pure-aluminum strips, while it is 1 W/m•K without. The enhanced quench propagation speed may be expressed by:

$$e = V_z / V_\phi = \{k_z / k_\phi\}^{1/2}$$

and one can expect the enhancement of the axial quench propagation by an order of magnitude.

As summary, the faster axial quench velocity can be expected in improvement of axial thermal conductivity. It may be realized by using pure aluminum strips placed along the coil axial direction on inner surface of the solenoid coil.

**2.5 Transparent vacuum vessel**

An outer vacuum vessel is a massive wall, because it need rigidity to withstand buckling mode forces due to external pressure. In order to minimize the material in the vacuum vessel as well as in the cold mass, a brazed honeycomb vacuum vessel has been investigated for SDC solenoid at SSC [21][22]. A major feature of honeycomb plate is the high stiffness, in spite of its light weight. Further, brazed honeycomb panel have the possibility for welding and high reliability due to the fact that no epoxy resin is used.

A prototype brazed honeycomb vacuum vessel was fabricated in order to ensure the honeycomb vacuum vessel design and to establish the fabrication method. A fabrication steps of a brazed honeycomb vacuum vessel is shown in Fig. 2.6. As described in Fig. 2.6, at bending process, a flat honeycomb plate can be accurately bent by the concept of 4 point bending method.

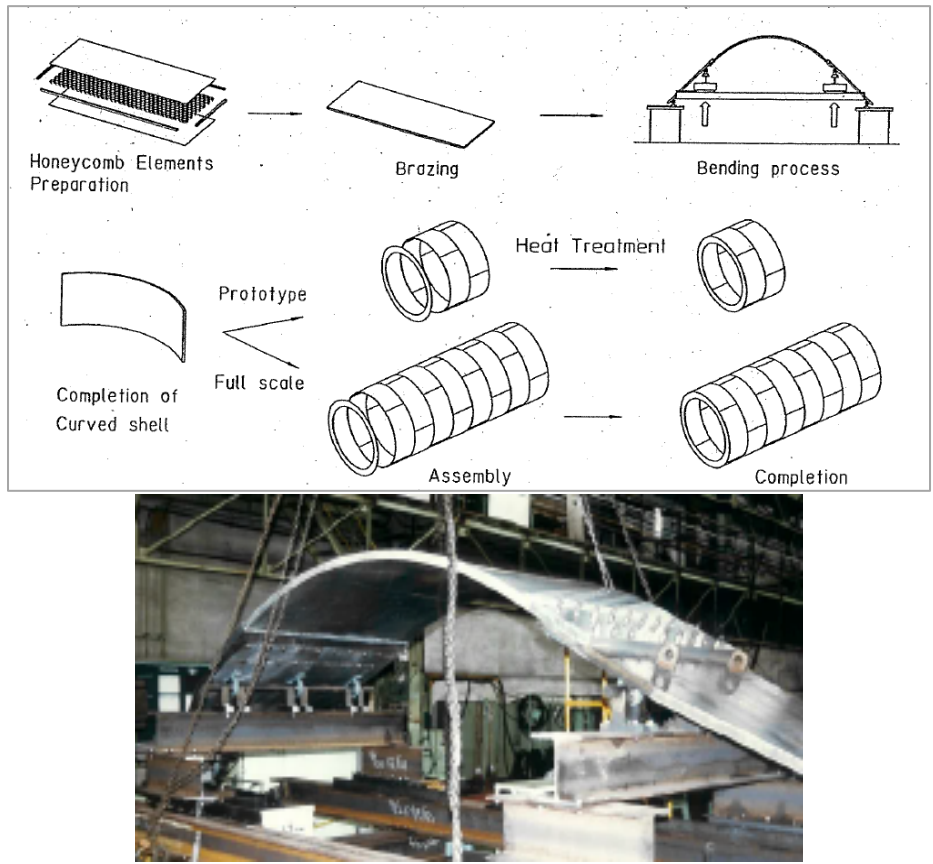

**Fig. 2.6** Fabrication steps of a brazed honeycomb vacuum vessel.

An alternative transparent vacuum vessel can be composed of aluminum isogrid shell. Aluminum isogrid shells are typically fabricated in steps: the grid pattern is first CNC machined in flat plates, then the flat plates are formed on a press brake into cylindrical sections which are welded to make up the shell, as shown in Fig. 2.7 [7][21].

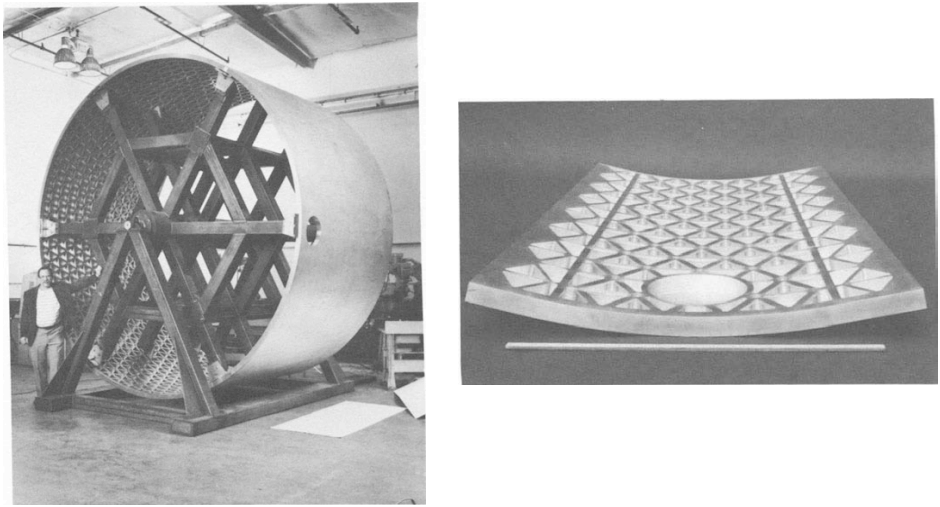

**Fig. 2.7** Welded isogrid shell on assembly fixture (left), and isogrid test panel with a hall for port.

**Table 2.2** summarizes the transparency comparison with solid, Al-honeycomb, and isogrid outer-vacuum vessel/wall, evaluated in the SSC-SDC detector solenoid R&D work [12]. The Al-honeycomb out vacuum wall may realize the best transparency under a normalized safety condition.

**Table 2.2** Comparisons of solid, isogrid and honeycomb outer vacuum vessel/wall evaluated for the SSC-SDC detector solenoid design.

| Type | Unit | Solid | Isogrid | Honeycomb |
|---|---|---|---|---|
| Al alloy | | 5083 | 5083-H32 | 6951/4045-T6 |
| # shells to be assembled | | 12 | 12 | 21 |
| Physical wall-thickness | mm | 27 | 46 | 46 |
| Skin wall-thickness | mm | (27) | 4.0 | 3.0+3.0 |
| Effective thickness (averaged) | mm | 27 | 11 | 7 |
| Weigh reduction ratio | | 1 | 0.4 | 0.26 |
| Radiation thickness | $X_0$ | 0.303 | 0.123 | 0.079 |

Although an isogrid shell will not be as thin as a honeycomb shell, it may have various advantages with the well-understood manufacturing process and easer arrangements to adapt cryostat parts, for example support structure bases or holes for cryogenic ports. Making good use of each advantage, isogrid and honeycomb technology may be well harmonized to compose a light weight and transparent vacuum vessel.

Further effort for an ultimately transparent vacuum vessel/wall has been made by using plastic material such as Carbon/Glass-Fiber-Reinforced-Plastic (CFRP/GFRP). An effort was the CFRP outer vacuum vessel for the TRISTAN-VENUS detector solenoid [7]. The effort has been progressed according to the reinforced plastic technology advances, and the technology is considered for future detector solenoids [24][25].

## 3. Future prospects for detector solenoid technology

The detector solenoids design study is in progress for future big projects in Japan and Europe, that is, ILC, FCC and CLIC. The proposed design parameters for each solenoid are summarized in Table 3.1

Table 3.1 The proposed design parameters for the detector solenoids for future projects

| Projects | Magnet | Bc (T) | InnerR (m) | Length (m) | E/M (kJ/kg) | Stored Energy (GJ) |
|---|---|---|---|---|---|---|
| **FCC-ee** | IDEA | 2 | 2.24 | 5.8 | 14 | 0.17 |
|  | CLD | 2 | 4.02 | 7.2 | 12 | 0.6 |
| **FCC-hh** |  | 4 | 5 | 20 | 11.9 | 13.8 |
| **CLIC** |  | 4 | 3.65 | 7.8 | 13 | 2.3 |
| **ILC** | ILD | 4 | 3.6 | 7.35 | 13 | 2.3 |
|  | SID | 5 | 2.5 | 5 | 12 | 1.4 |

The main development item for the future detector solenoid is Al stabilized superconducting cable with both higher strength and keeping high RRR, and most likely solution is the combination of technologies used in the conductors for ATLAS-CS and CMS. The approach adopted in ATLAS-CS conductor is the homogeneous reinforcement of the aluminum stabilizer, that is, to dope Ni into Al stabilizer and to apply cold-work hardening simultaneously. Another approach used in CMS conductor is the reinforcement by using the hybrid configuration with pure-aluminum stabilized conductor and high strength aluminum alloy, which are mechanically bonded by electron beam welding. By combining these two approaches, the (0.2%) yield strength of more than 300 MPa might be expected as shown in Fig. 3.1 [26]. Various conductors are designed in the future projects described in the next section.

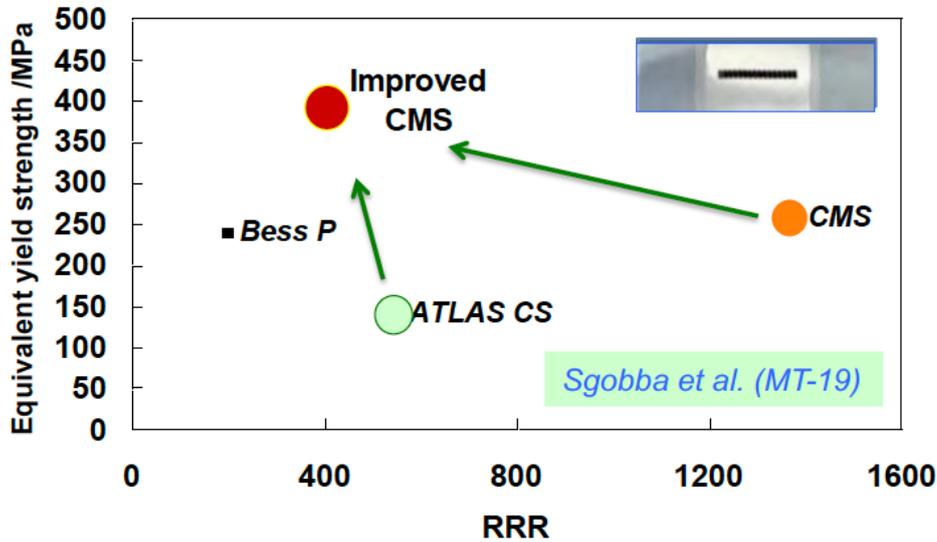

**Fig. 3.1** Yield strength and RRR which were realized in ATLAS-CS and CMS. Expected parameters by combining two technologies is also plotted.

The other magnet fabrication technologies will be also progressed with the advanced Al-stabilized conductor, and the technologies based on the present ones will be used adaptively for each project; inner coil winding technique inside the outer shell, indirect cooling to reduce materials used in magnet structure, and utilize pure aluminum strips as temperature equalizer in the steady operation and fast quench propagator for the safe quench protection.

The vacuum vessel design is one of the important studies especially for the detector solenoid requiring the transparency for the charged particle passing through. The vessel design with rigidity, light-weight and high transparency will be required. The materials used for the vacuum vessel is also interesting topic. The material technology is rapidly developed and options such as plastic composites are of interest [25].

An effort in design and studies for the use of high temperature superconductor are also requested to reach the increasing demand on sustainability in reducing power consumption of our superconductor detector magnets, even for minimizing the cryogenics power consumption.

A concern for future large-scale magnet development is the inheritance of present technologies as well as the technological breakthrough. The technologies described in the section 2 have been used successfully for the manufacturing of many superconducting detector magnets in the past four decades thanks to the technical and scientific competencies developed, with many regular breakthroughs. However, the large-scale detector magnets with Al-stabilized conductors have been not fabricated since after the success of CMS and ATLAS-CS in LHC. Today complementary efforts are needed to resume again an equivalent level of expertise, to continue the effort on research and to develop these technologies and apply them to each future detector magnet project. Especially, it is mandatory for the development of Al-stabilized conductor to get the industry involved. A collaborative framework between institutes and industries is needed, and pre-industrialization programs will be necessary to adapt the technologies to the specific needs of the new detector magnets, and to validate that the required quality and performances can be reached.

In addition, quality control and assurance are necessary from a long-term perspective. The recent maintenances done at CERN on the LHC detectors during the long shutdown 1 and long shutdown 2 revealed that an effort will also have to be made on quality assurance during the design and procurement phases together with the industrial suppliers for each sub-system to reach the necessary standards for detector magnets that will be operated in high energy physics experiments for the coming decades.

In the discussion above, a major topic of discussion is a solenoid-shaped magnet as a detector solenoid. Other types of magnets, such as split coil and saddle shape coil, could be candidates depending on the requirements from a physics viewpoint [27]. The continuous effort of coil design study is important to make an magnet design optimized for experimental goals.

The detector solenoid for mid-scale experiment sometimes uses a conventional Cu-stabilized Nb-Ti conductor. The conventional solenoids for such experiments do not need the development of unique conductor, but instead, precise control of magnetic field distribution might be required. The development efforts are on-going in terms of the magnetic field design technology with high precision simulation, the coil fabrication technology to achieve the design requirement and the control method of magnetic field distribution [28].

## 4. Future Projects

### 4.1 Detectors for high energy physics

#### 4.1.1 FCC-ee

The Future Circular Collider (FCC) is a project proposed to start after the Large Hadron Collider (LHC) at the Centre for Nuclear Research (CERN) [29]-[31]. FCC is a circular particle accelerator with a circumference of 100 km and a proposed center-of-mass energy of 100 TeV in case of proton-proton collisions, which would be FCC-hh. However, the first stage of the FCC is foreseen to be the FCC electron-positron collider (FCC-ee) that can be used to study the electroweak sector with unprecedented accuracy [32].

At the moment of writing this, there are three detector designs for FCC-ee: the Innovative Detector for Electron-positron Accelerators (IDEA, [33]), the CLIC-Like Detector (CLD, [34]) and a design comparable to the IDEA detector that remains to be named. Each of these three designs includes a superconducting solenoid that produces a 2 T magnetic field in the center of each detector. The CLD magnet is positioned outside of the calorimeter volume while the two other solenoids are situated inside the calorimeter barrels just after the tracking detectors. Since the latest (nameless) detector design and its solenoid are still work in process the following paragraphs will be focused on the IDEA and CLD magnets.

FigureFig. 4.1 shows the magnetic flux density as a function of the location in the axisymmetric plane of the CLD and the IDEA detector designs, where the locations of the different sub-detectors are indicated as well.

FigureFig. 4.1 also highlights the most important difference between the two detector designs, which is the location of the solenoid with respect to the calorimeters. Since the IDEA magnet is inside the calorimeter volume there are strict requirements on the particle transparency that needs to be lower than 1 $X_0$. The concept of the solenoid of the IDEA is similar to the ATLAS Central Solenoid [35] and whereas the CLD magnet is similar to the CMS solenoid [36]. However, in terms stored energy and free-bore diameter there are some important differences.

The free-bore diameter of the IDEA solenoid is 4.2 m and is almost two times bigger than the free bore of the ATLAS CS. This also means that the IDEA magnet has around four times the stored magnetic energy of the ATLAS CS at 170 MJ. The free-bore diameter of CMS is 6 m with a stored energy of 2.6 GJ. The CLD design has a larger free bore of 7.2 m and its stored energy is 600 MJ. This results in the following design parameters for the IDEA and the CLD summarized in Table 4.1 [37].

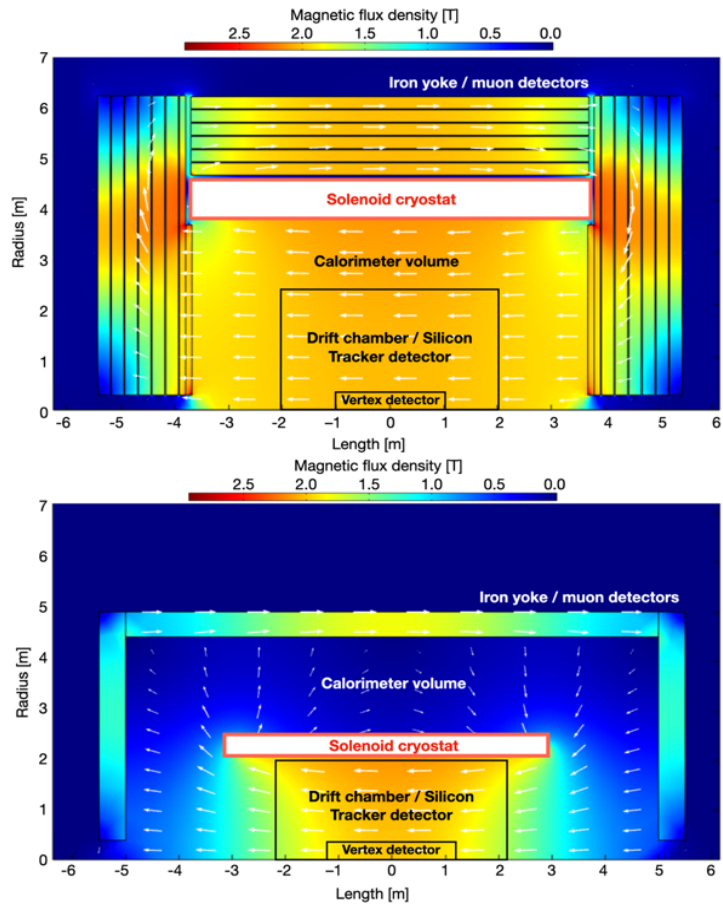

**Fig. 4.1** Magnetic flux density as a function of the location in the axisymmetric plane of the IDEA (TOP) and the CLD (Bottom) detector designs

To highlight the challenges of the IDEA and CLD designs the energy density, i.e. the stored magnetic energy divided by the cold mass weight, can be used as an indication. The ATLAS CS has a maximum energy density of 7.0 kJ/kg during nominal operation (with a demonstrated maximum of 8.1 kJ/kg) while for the IDEA magnet the energy density would be twice as high at 14 kJ/kg. Similarly, the CMS energy density is 11 kJ/kg and the CLD magnet has an energy density of 12 kJ/kg. For reference, the highest energy density ever produced in a detector magnet was 13 kJ/kg in the BESS experiment [38]. These high energy densities mean that quench protection needs to be both active and passive in case of a quench detection failure, as well as redundant. The large free-bores in combination with the high stored energies translate to strong requirements on mechanical support and strength of the materials used, especially in the case of the IDEA magnet that is not only very large but the cold mass is extremely thin at only 53 mm [37].

**Table 4.1** Design parameters of the superconducting solenoids for the IDEA and CLD detector concepts at the FCC-ee.

| Property | IDEA | CLD | Unit |
|---|---|---|---|
| Conductor | | | |
| Conductor material | NbTi/Cu in Al/Ni cladding | | |
| Conductor height | 36 | 36 | mm |
| Conductor width | 10 | 22 | mm |
| Turn-to-turn insulation | 1 | 1 | mm |
| Number of strands | 30 | 26 | |
| Strand diameter | 1.1 | | mm |
| Cu:SC fraction | 1 : 1 | | |
| Operating current | 20 | | kA |
| Operating temperature | 4.5 | | K |
| Coil | | | |
| Inner radius | 2.235 | 4.02 | m |
| Length | 5.8 | 7.2 | m |
| Weight | 12.5 | 49.5 | t |
| Number of turns x layers | 530 x 1 | 300 x 2 | |
| Support cylinder thickness | 12 | 25 | mm |
| Total coil thickness | 53 | 102 | mm |
| Central field | 2 | | T |
| Stored magnetic energy | 170 | 600 | MJ |
| Energy density | 14 | 12 | kJ/kg |

Starting with the mechanical support, both the IDEA and CLD include an aluminum 5083 support cylinder surrounding the coil windings [37]. The yield strength of Al5083 is higher than 209 MPa at 4.2 K [39]. The aluminum stabilized conductor proposed for these magnets has a yield strength of 147 MPa at 4.2 K when taking the Nb-Ti into account [35]. The conductor will be glued on the inside of the support cylinder with an epoxy resin type adhesive. These can have a shear strength of up to 76.8 MPa at 77 K depending on the type of resin used [40].

Initial mechanical simulations reveal that the IDEA magnet has a peak hoop stress of 105 MPa and for CLD this is 75 MPa [37]. Both these values are within the elastic regime of the conductor and the Al5083 support cylinder. The peak tensile strains for IDEA and CLD are 0.13 % and 0.11 %, respectively. At the interface between the coil windings and the support cylinder the peak shear stresses are 0.5 MPa for IDEA and 0.24 MPa for CLD, which is well within the maximum shear stress that adhesives can tolerate.

Large scale detector magnets like the IDEA and CLD magnet are only built once and there are usually no representative prototypes. Furthermore, because they are in between different particle detectors it is not possible to replace them in they break down. Therefore, it is crucial to have redundant quench protection, even allowing for the case where quench detection fails. An example of the energy extraction layout during a quench is shown in Fig. 4.2. Quench simulations for the CLD magnet showed that with an extraction resistor or Run-Down Unit (RDU) the peak temperature after a quench is 60 K and the magnet is fully discharged after 600 s [37].

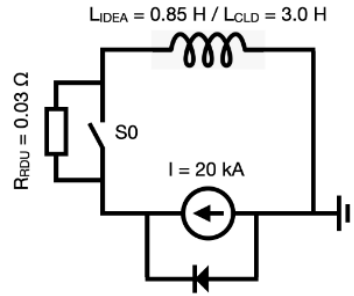

**Fig. 4.2** Example of the energy extraction circuit during quench.

The IDEA solenoid has a larger energy density, a smaller wall thickness and higher stresses than the CLD magnet [37]. Therefore, a fully three-dimensional simulation of the IDEA magnet was used to study quench protection measures. In addition to an RDU, quench heaters and high purity aluminum (RRR = 3000) quench propagation strips (QP strips) were studied. These QP strips are found in the ATLAS CS as well [35].

The results of the 3D simulations for different quench scenarios are shown in Fig. 4.3 [37]. The cases with an RDU have the lowest hot-spot temperature equal to 65 K. In all other quench scenarios without a protection resistor, it is seen that the QP strips have a large effect on the peak temperature. In some cases with QP strips, the peak temperature is decreased by more than 100 K compared to when the aluminum strips were not present. In addition, in cases with QP strips sixteen turns quench before the quench is detected and validated while in the case without QP strips only eleven turns quench before the quench is detected. This means that the quench propagation strips have a large benefit in terms of quench protection. A benefit of the quench protection strips is that they are fully passive and they even work in case the quench was not detected by the safety systems.

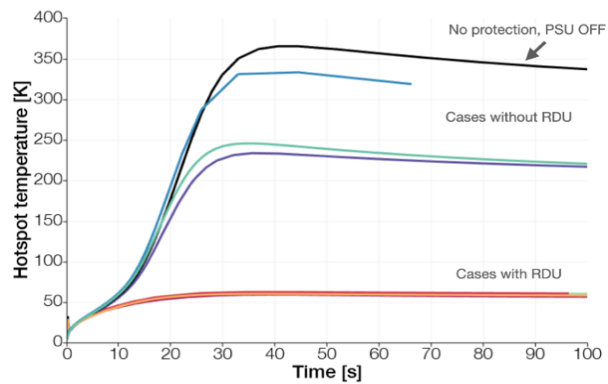

**Fig. 4.3** Results of the 3D simulations for different quench scenarios

In the previous paragraphs preliminary studies on the FCC-ee detector magnets are described. One of the future problems for detector magnets is the availability of the aluminum stabilized conductor that is also needed for detector magnets of FCC-ee. In addition, the mechanical support, a more detailed quench analysis, the service lines, cryogenics and the cryostat, magnet operation and control are among other topics are still need to be researched and developed in the coming years to enable the construction of these very challenging superconducting solenoids.

**4.1.2 FCC-hh**

The FCC-hh project foresees a significant 7-fold increase in the particle collision energy with respect to the LHC, and to measure the momentum of the highly energetic collision products with sufficient resolution much more powerful detector magnets are needed as well [41].

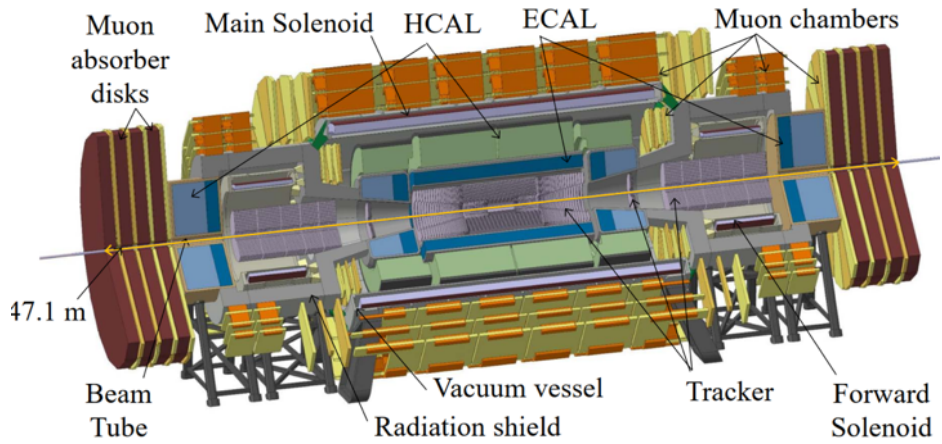

**Fig. 4.4** Proposed FCC-hh detector base-line layout

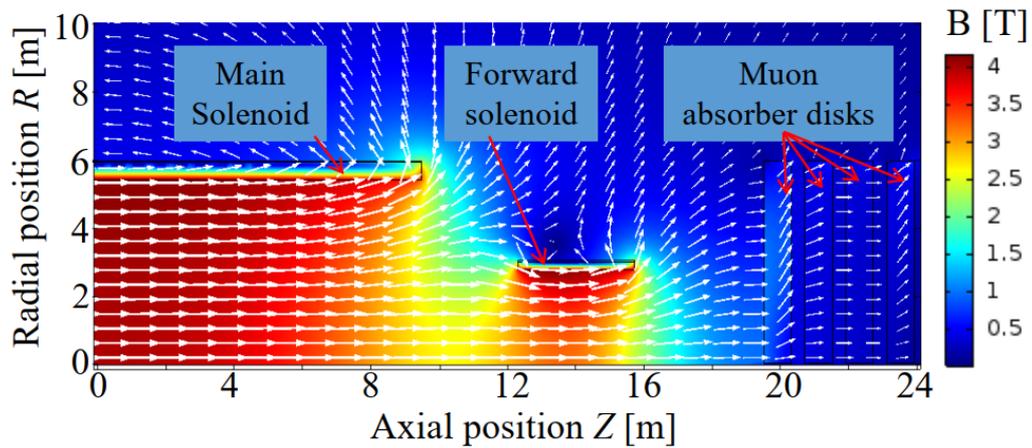

**Fig. 4.5** Magnetic field map of the proposed FCC-hh baseline detector magnet configuration

For this purpose, the FCC-hh detector magnet layout (Fig. 4.4) was proposed, featuring three powerful superconducting solenoids (Fig. 4.4 and Fig. 4.5) each generating 4 T in their bore [42]. Here the central solenoid features are a 10 meter free bore diameter and a length of 20 meters, whereas the forward solenoids features are a cold mass length of 3.4 meters and a free bore diameter of about 5 meters.

This detector layout has some similarity to the CMS detector, featuring a tracker, electro-magnetic calorimeter (E-CAL), and a hadron calorimeter (H-CAL) in the bore of the magnet, and muon chambers on the outside of the magnet. The muon chambers utilize the magnetic return flux generated by the main solenoid for the purpose of muon tagging. Unlike CMS, the FCC-hh detector does not feature iron yokes.

The unique combination of a main and forward solenoids is proposed to enhance the momentum resolution for particles travelling nearly parallel to the bore tube, and for this purpose trackers and calorimeters are located both in the main and the forward solenoids (Fig. 4.4). Due to the close proximity of the main and forward solenoids, the forward solenoids are each exposed to an net attractive force of 60 MN towards the main solenoid. This force is transferred to the solenoid vacuum vessels through reinforced tie rods, and the force is subsequently transferred to the main solenoid through the vacuum vessels and additional support structure.

**Table 4.2** Overview of various detector magnet properties for the FCC-hh proposed detector magnet concept

| Property | Value |
|---|---|
| Total stored energy [GJ] | 13.8 |
| Operating current [kA] | 30.0 |
| Combined inductance [H] | 30.7 |
| Inductance main solenoid [H] | 27.7 |
| Inductance forward solenoid [H] | 0.93 |
| Mutual inductance, main-to-forward [H] | 0.29 |
| Mutual inductance, forward-to-forward [H] | 0.001 |
| Cold mass main solenoid [t] | 1070 |
| Cold mass forward solenoid [t] | 48 |
| Vacuum vessel main solenoid [t] | 875 |
| Vacuum vessel forward solenoid [t] | 32 |
| Average energy density [kJ/kg] | 11.85 |
| Minimum shaft diameter [m] | 13 |

Table 4.2 shows various properties of the proposed superconducting detector magnets for FCC-hh. The bore magnetic field of 4 T and the energy density of 11.85 kJ/kg are a bit higher than CMS, while the total stored magnetic energy of 13.8 GJ is more than five times higher. The peak Von Mises stresses in both the main and the forward solenoid are at 100 MPa under nominal conditions which illustrates that, similar to CMS and the ATLAS Central Solenoid, a reinforced conductor is required to handle the Lorenz forces.

Fig. 4.6 shows the proposed conductor geometry. The main and forward solenoids feature 8 and 6 layers respectively. The conductors comprise Nb-Ti/Cu Rutherford cables surrounded by nickel-doped aluminum stabilizer. The Nb-Ti/Cu Rutherford cables feature 40 strands with a diameter of 1.5 mm, a Cu:non-Cu ratio of 1:1, and a current sharing temperature of 6.5 K. The operating current is 30 kA.

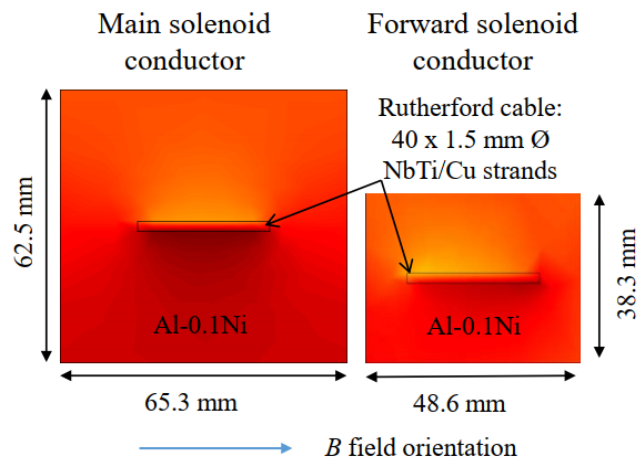

**Fig. 4.6** Proposed conductor geometry, featuring nickel-doped aluminum-stabilized Nb-Ti/Cu Rutherford cables

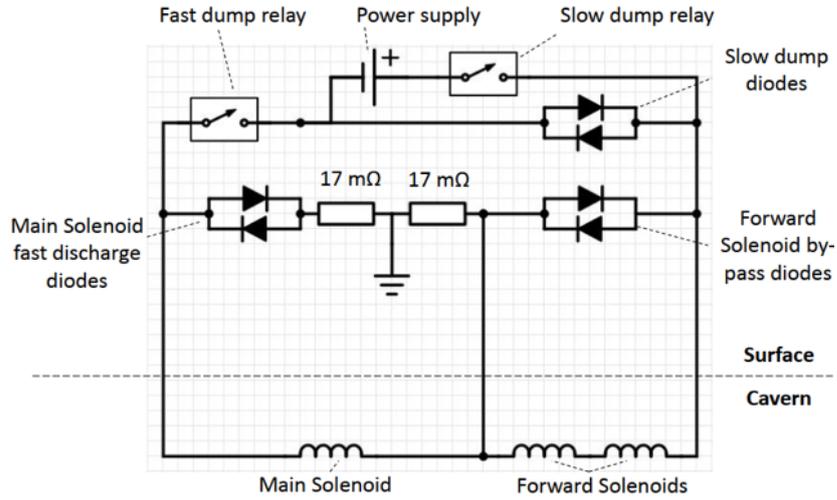

**Fig. 4.7** Proposed circuit layout for powering and discharging the FCC-hh detector solenoids

To power the solenoids, and initiate slow and fast discharges as needed, a combined circuit is proposed (**Fig. 4.7**). The solenoids are powered in series so that a single power supply and slow dump circuit are sufficient to charge and discharge the solenoids. In case of a quench, the different operating current densities in the main and forward solenoids necessitates a current decoupling of these two different magnet types, and therefore the two magnet types each feature their own fast-discharge dump system comprising diodes and resistors. Moreover, in case of a quench normal zones are induced in various spots on the solenoids through quench heaters to quickly bring them to normal state, thus avoiding strong temperature gradients even under fault conditions where the fast dump units fail to discharge the magnets. This proposed protection scheme thus provides redundant protection. The calculated peak hot-spot temperature is well below 100 K under nominal conditions (**Fig. 4.8**)

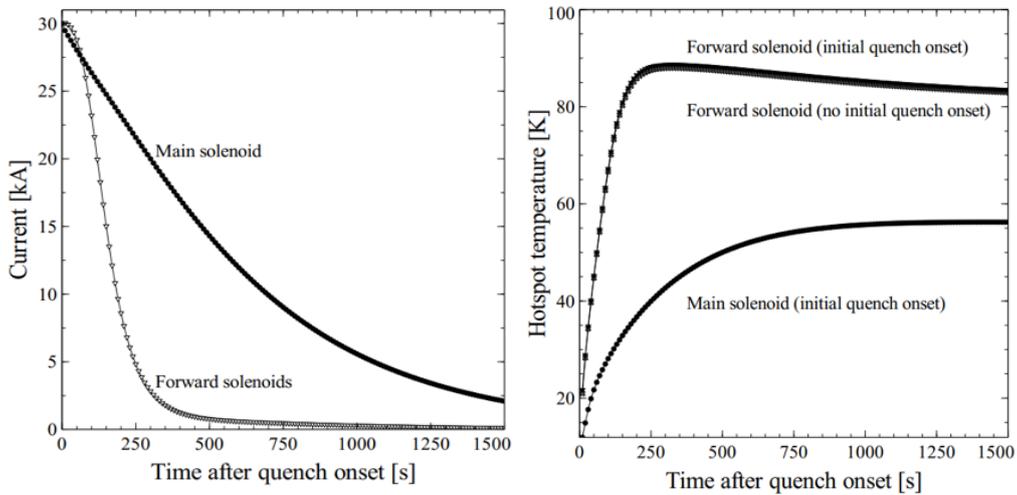

**Fig. 4.8** Simulated current discharge and hot-spot temperature development in case of a quench

In summary, a conceptual design of superconducting detector solenoids was previously developed for the FCC-hh detector [42]. The bore magnetic field, energy density, and mechanical stress are similar to CMS, but the stored magnetic energy and overall cold mass is more than five times larger. This necessitates the use of a mechanically reinforced aluminum-stabilized conductors to handle the Lorentz forces. A particular feature of the FCC-hh detector is the combination of three solenoids which gives superior performance for particle tracking albeit at the cost of additional complexity and a large net force of 60 MN on each of the two forward solenoids.

**References**

[41] "FCC-hh: The Hadron Collider", *EUR. Phys. J. Special Topics* 228, p. 755-1107 (2019)
[42] M. Mentink, H. Silva, A. Dudarev, E. Bielert, V. Klyukhin, B. Cure, H. Gerwig, A. Gaddi, C. Berriaud, U. Wagner, and H. ten Kate, "Evolution of the Conceptual FCC-hh Baseline Detector Magnet Design", *IEEE Trans. On Appl. Supercond.* 28, p. 4002710 (2018)

### 4.1.3 CLIC

The Compact Linear Collider (CLIC) detector project Collaboration [43][44] intends to build the CLICdet, the new CLIC detector model that has been updated after the CLIC Conceptual Design Report (CDR) was published [34], a detector with a 4-T solenoid magnet, operated at the three stages of the CLIC accelerator phases with center-of-mass energies of 380 GeV, 1.5 TeV, and 3 TeV. A view of the detector cross section in the interaction region is shown in Fig. 4.9.

The feasibility of having a dual beam delivery system serving two interaction regions for the Compact Linear Collider has recently been studied and looks promising for physics programs [45]. At this present stage, no distinction has been made on the magnets for these two detectors. It is the CLICdet baseline magnet design that has been considered with two crossing angles to check the physics feasibility in the interaction regions with the CLIC dual beam delivery system.

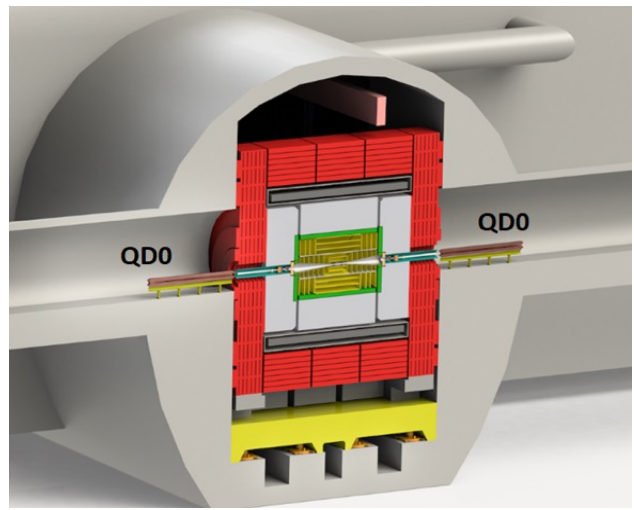

**Fig. 4.9** Vertical cut through of the CLICdet detector model showing the QD0 final focusing quadrupole positions outside the detector.

The CLICdet magnet design is based on the designs and manufacturing breakthroughs of the CMS solenoid [46] and the Atlas Central Solenoid (CS) [47]. The CLICdet magnet is a 4-T solenoid with 4 layers of superconductor. The main parameters are indicated in Table 4.3. The coil is shorter compared to the CMS solenoid, but it has a larger radius, which gives a total magnetic flux that is about 45% higher. A 3D view is presented in Fig. 4.10.

The superconducting conductor comprises a Rutherford cable composed of 32 Nb-Ti strands that is stabilized with an aluminum sheath. This is quite comparable to CMS, with a bit larger conductor. The conductor will have to be reinforced to accommodate the large magnetic forces applied to the winding. Both options of ATLAS CS and CMS for a reinforced conductor are considered at this stage, respectively the structural cold worked Ni-doped aluminum stabilizer [18] and the Electron-Beam welded aluminum alloy reinforcement [39]. For both options the Rutherford Nb-Ti superconducting cable will be co-extruded with the high Residual Resistivity Ratio (RRR) aluminum stabilizer. Feasibility programs were led on the structural doped aluminum with a large cross section [49]. Other studies of reinforcement proposed ways to increase the mechanical properties of the conductor, based on the ATLAS CS and CMS reinforcement concepts [26][50].

**Table 4.3** CLICdet magnet parameters

| Property | Value |
|---|---|
| Magnetic field at IP [T] | 4 |
| Inductance [H] | 12 |
| Nominal current [kA] | 20 |
| Stored energy [GJ] | 2.3 |
| Average energy density [kJ/kg] | 13 |
| SC cable number of NbTi strands | 32 |
| Conductor cross section [mm$^2$] | 83 x 20 |
| Coil inner radius [mm] | 3650 |
| Coil length [mm] | 7800 |

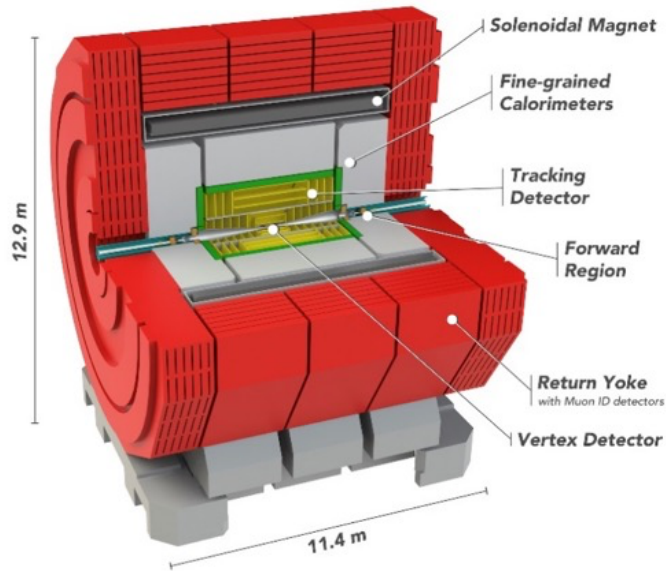

**Fig. 4.10** The 3-D view of the CLICdet model with external dimensions and cut-out showing the main sub-systems.

The magnet will be inside a cryostat supported by the central barrel yoke, similar to CMS. The CLICdet coil will be built using the inner winding technique inside a 50-mm thick external mandrel serving as structural external wall, as support for the liquid helium cooling circuit, and as fixation for the supporting tie rods. Similar to CMS, the external mandrel is made from an aluminum alloy of the 5083 grade. It is also used for the protection of the coil against the quench acting as a so-called quench-back cylinder. Aluminum thermal strips shall be used to ensure a good temperature uniformity in the cold mass, during cool down, operation at 4K, warm-up and for quench protection. Quench heaters are also proposed. The coil will be built from 3 modules with splices integrated on the low field region on the outer radius of the coil, similar to the CMS coil [51][52]. The vacuum impregnation technique is considered. Heat radiation shields, together with an indirectly cooling with boiling helium at 1.2 bar and in thermosiphon mode will allow the operation of the magnet at 4K.

The CLICdet detector has an iron yoke used to confine the magnetic flux and take benefit of it for the muon detectors installed in between the iron layers. A set of 4 end coils are attached on each end cap of the detector. The cross section of the magnet is given in Fig. 4.11. The end coils are used both to limit the magnetic field in the machine detector interface region, as seen in Fig. 4.12, in particular on the QD0 final focusing quadrupoles located just in front of the end caps, and to limit the stray field outside the detector to 16mT at a radial distance of 15m in the service cavern that is used for detector maintenance and where detector services, powering systems and cryogenics are located (Fig. 4.13). The use of these ring coils also contributes to limit the amount of iron in the yoke. It was proposed in the CLIC CDR [53] to build these end coils with normal conducting windings operated at room temperature and water cooled, but we see here a good application of more sustainable solutions in order to limit the heat losses due to the dissipated power by using high temperature superconductor coils, connected in series.

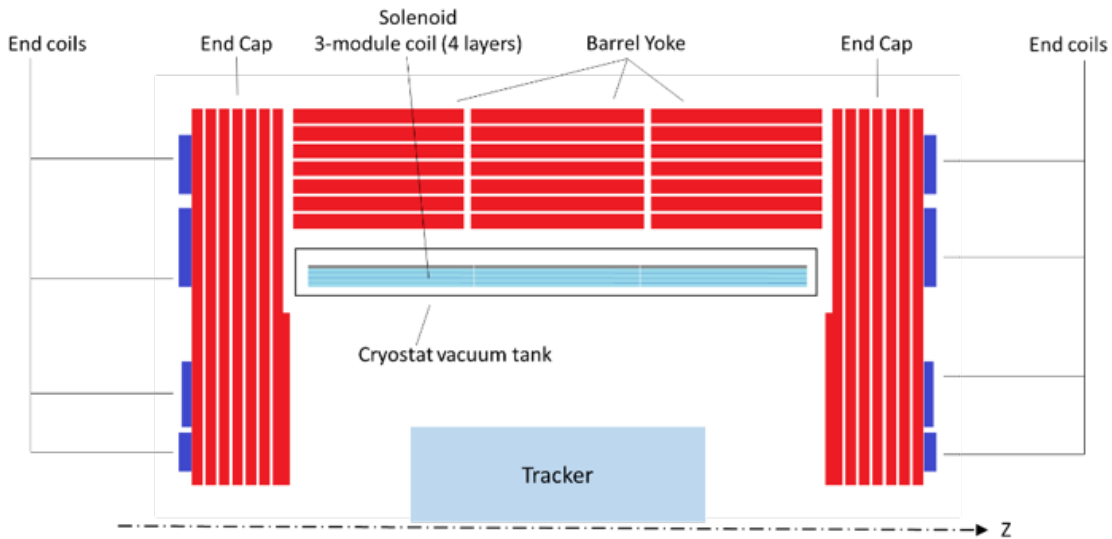

**Fig. 4.11** Schematic RZ view of the CLICdet magnet. Only one half of the magnet section along its axis is shown and calorimeters as well as other detector details are not represented.

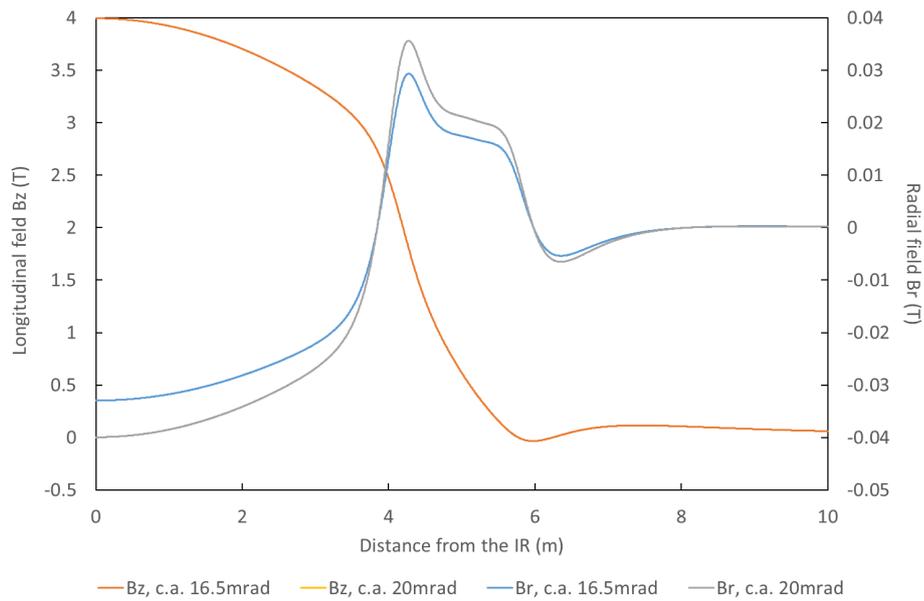

**Fig. 4.12** The radial Br and the longitudinal Bz magnetic fields from the IR up to 10m along the beam axis.

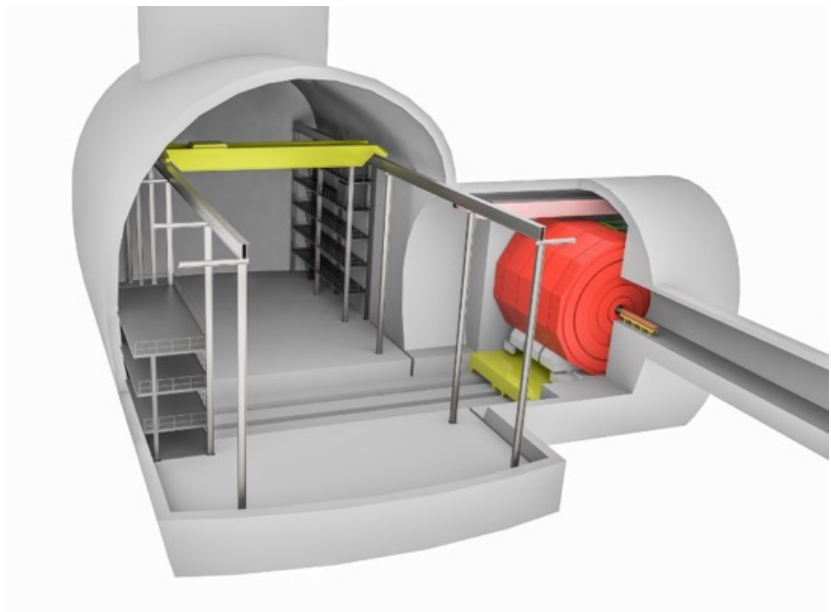

**Fig. 4.13** Schematic view of the infrastructure layout in the experimental underground cavern.

Several technology R&D and pre-industrialization programs will be needed before launching the manufacturing of such a magnet.

Power leads [54][55] and superconducting busbars [56][57] are typical applications that can be developed using high temperature superconductors. Other developments can also be performed for powering DC converters and dumping circuit. Dedicated studies applied to the detector magnet applications will be needed.

Specific equipment and tooling for the conductor and coil manufacturing (cabling machine, cable brushing and preheating, co-extrusion dies, continuous welding, winding, etc.) and quality control and measurement devices (e.g. continuous quality control of the superconductor, impregnation quality, field mapping) will have to be adapted or re-developed specifically ahead of this project during a pre-industrialization program.

### 4.1.4 ILD

The design parameters of the magnet for the International Large Detector (ILD), as shown in Fig. 4.14, feature a central field of up to 4 T, in a volume of about 275 m$^3$ (useful diameter 6.88 m over a length of 7.35 m) and its conceptual design has been undertaken by CEA, DESY and CERN [58][59][60]. The ILD magnet design is very similar to the one of CMS, except for its geometrical dimensions, and the presence of the anti-DID. Consequently, many technical solutions successfully used for CMS are proposed for the design of the ILD magnet. The winding radius (3.615 m) of the ILD has larger figure-of-merit of 58.4 T$^2$m than the value of CMS (47.2 T$^2$m) due to its larger size. Similarly to CMS, a 4-layer coil is retained, with a nominal current in the range of 20 kA, so the conductor for the ILD magnet has larger cross section of 74.3 x 22.8 mm$^2$. It consists of a superconducting Rutherford cable, clad in a stabilizer and mechanically reinforced. Two solutions are considered for the reinforcement. The first option is a micro-alloyed material such as the ATLAS central solenoid [14], which acts both as a stabilizer and a mechanical reinforcement. A R&D program on the Al-0.1wt%Ni stabilizer has been launched at CERN and is underway to demonstrate the feasibility of producing a large conductor cross section with this material. The second option is a CMS-type conductor with two aluminum alloy profiles welded by electron beam to the central conductor stabilized with high purity aluminum. These two options are shown in Fig. 4.15, together with the actual CMS conductor for comparison.

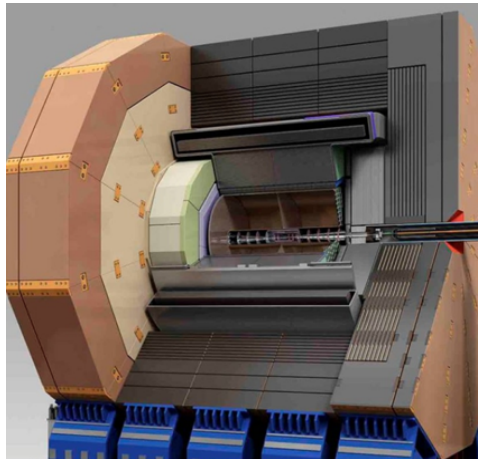

**Fig. 4.14** Configuration of ILD

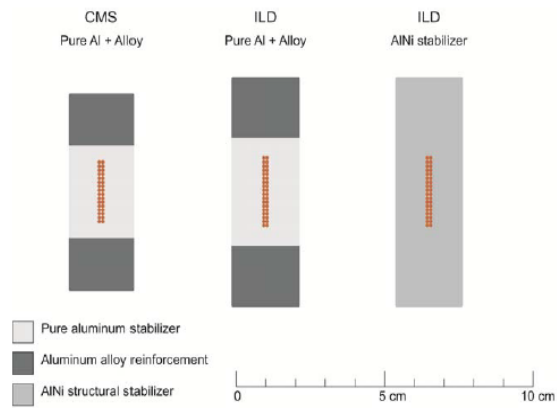

**Fig. 4.15** Two options of ILD conductor composition.

### 4.1.5 SID

A conceptual design study for a 5 Tesla superconducting solenoid for the Silicon Detector (SiD as shown in Fig. 4.16) of the International Linear Collider (ILC) has been undertaken in FNAL [61][62][63]. The solenoid has a clear bore with 5.0 m in diameter and 5.0 m in length, where 5 T magnetic field is produced for inner detectors. Although the winding radius (2.65 m) of the SiD coil is smaller than that of the CMS (2.95 m), it has larger figure-of-merit of 62.5 $T^2m$ than the value of CMS (47.2 $T^2m$) due to its higher magnetic field. Utilizing the existing CMS magnet conductor as the starting point, a winding design has been proposed for the magnet as shown in Fig. 4.17. Finite element analysis shows the resulting magnetic stresses in the coil parts do not greatly extrapolate beyond those of CMS.

Major R&D subject for SiD magnet is its conductor required to sustain such large EMF in the coil. In "SiD Letter of Intent" described in 2009 [63], a more advanced and most likely cheaper conductor was proposed. It is based on high purity aluminum alloys, such as Al-0.1%Ni, which were used in the ATLAS central solenoid. FigureFig. 4.18 shows a cross sections of the CMS conductor and of an advanced SiD conductor design using Al-0.1%Ni and novel internal high strength stainless steel reinforcement. Other conductor stabilizer possibilities are also under consideration and study. These include TiB2 grain refinement aluminum matrix composites, and cold working via the equal area angle extrusion process

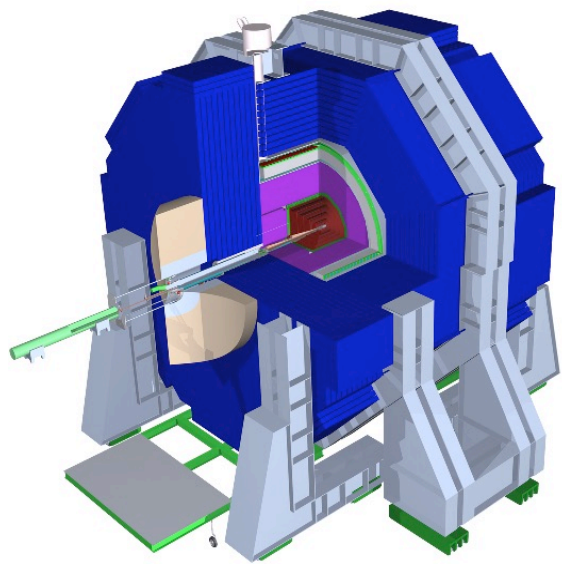

**Fig. 4.16** Overview of the SiD

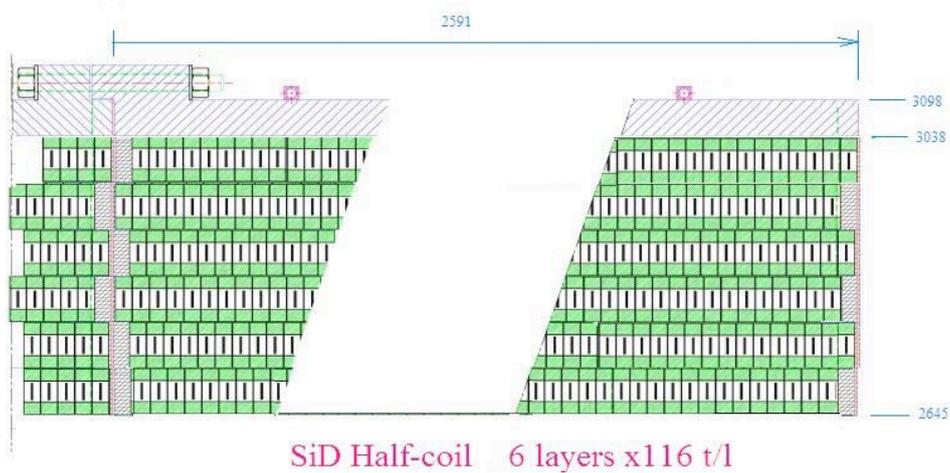

**Fig. 4.17** The SiD coil winding design. The cooling pipes are shown welded to the outer support cylinder; the two modules are joined by bolting at the median plane.

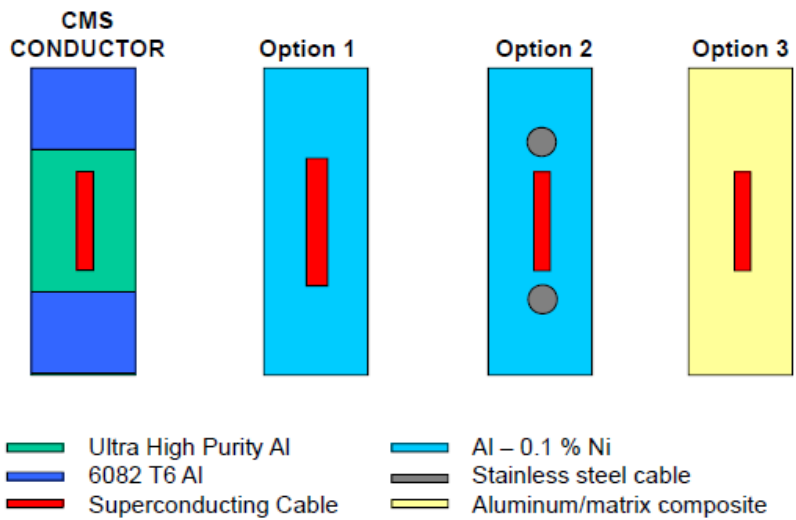

**Fig. 4.18** Cross sections of CMS conductor and proposed SiD conductors.

**References**

[61] Updating the SiD detector concept. https://arxiv.org/abs/2110.09965.
[62] R. P. Smith and R. H. Wands, "A five tesla solenoid with detector integrated dipole for the silicon detector at the international linear collider," *IEEE Trans. Appl. Supercond.*, Vol 16, No. 2, 2006.
[63] "SiD Letter of Intent" edited by H. Aihara, P. Burrows, M. Oreglia, 2009. https://insprehep.net/literature/835665

**4.2 Detectors for secondary particle experiments**

**4.2.1 COMET**

The COMET experiment in J-PARC aims to explore the rare decay phenomenon of muons. Fig. 4.19 shows the overview of COMET Phase-I beam line. In order to transport the muons effectively, the superconducting solenoids are used throughout the muon beamline from the target

– 29 –

to generate pion, to the electron detectors, that is, the Pion Capture Solenoid (PCS), Muon Transport Solenoid (MTS), Bridge Solenoid (BS) and Detector Solenoid (DS). The magnetic field of all the magnets is required to be connected smoothly without large dips and peaks, otherwise the muons might be trapped or reflected. Fig. 4.20 shows the magnetic field distribution on the muon beam axis. The solenoids are aligned in such a way that the ripple of magnetic field would be less than 5 % of averaged field in each section. All the solenoids are covered by iron yokes, which are used as both radiation shield and magnetic flux return yoke.

Fig. 4.21 shows the 3D view of DS for COMET Phase-I experiment. The detector solenoid acts as a spectrometer of the electrons generated by the decay of muons, and all the electron detectors are installed inside the magnet bore. The detector solenoid is conventional solenoid wound by copper-stabilized Nb-Ti conductor cooled by GM cryocoolers. The total length of superconducting coil is 1.9 m, inner bore diameter of coil is 2.14 m. The superconducting coil consists of 14 coils of 170 mm in length and 8 mm in thickness. All the coils are connected in series, and the nominal current is 189 A to generate the central field of 1 Tesla.

The inductance of the detector solenoid is so large, 236 H, to extract the stored energy into the dump resistor for the quench protection, therefore, the quench protection by a passive heater is adopted. FigureFig. 4.22 shows the quench protection circuit of DS. Heater wire of 1.5 mm in diameter is wound on the outside of superconducting coils, and the heaters are connected in parallel with coils as shown in Fig. 4.22. When the quench is detected by the quench detection system, the power supply is cut off by the circuit breaker, and the magnet current go through the heaters. Thanks to the quick quench propagation by the heaters, the maximum temperature in the coil can be suppressed below 150 K during the quench.

The PCS is not a detector solenoid, but the technology of the detector magnet is adopted, such as, the superconducting cable stabilized with high purity aluminum [64]. The PCS contains the pion production target, and it is exposed to high radiation, meaning that large heat load is expected into the coils, calculated to be 228 W at maximum in the Phase-II experiment. In addition, conduction cooling scheme is applied in order to reduce the exposure of the liquid helium to direct radiation. The Nb-Ti with copper stabilizer based thick aluminum stabilized cable is used in the PCS; 15 mm in width, 4.7 mm in thickness and composition ratio of Al/Cu/Nb-Ti is 7.3/0.9/1.0. The magnets are cooled down by cooling pipes flowing two-phase liquid helium on the outer surface of the coil shell, and the pure aluminum strips are sandwiched between layers as shown in Fig. 4.23, to help the removal of radiation heat.

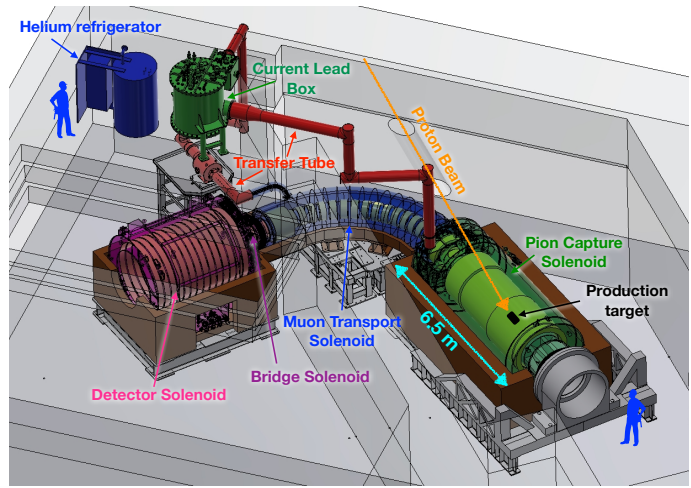

**Fig. 4.19** Overview of COMET beam line for Phase-I experiment.

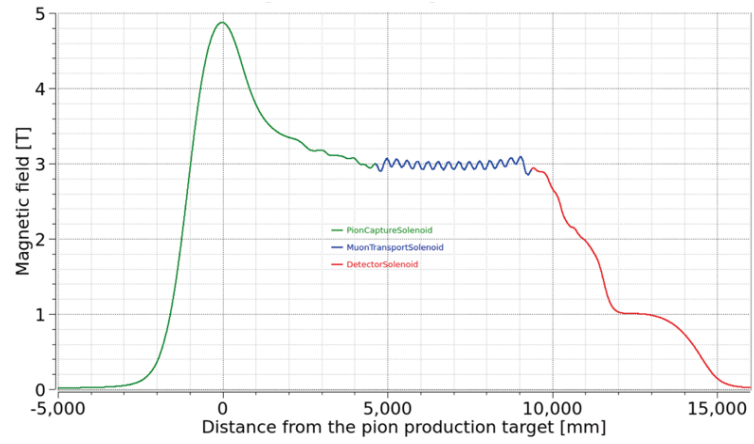

**Fig. 4.20** Magnetic field distribution of COMET beam line for Phase-I experiment.

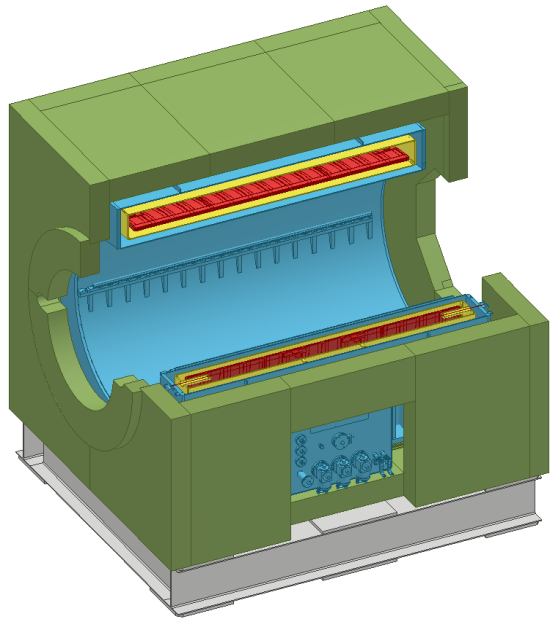

**Fig. 4.21** COMET DS magnet.

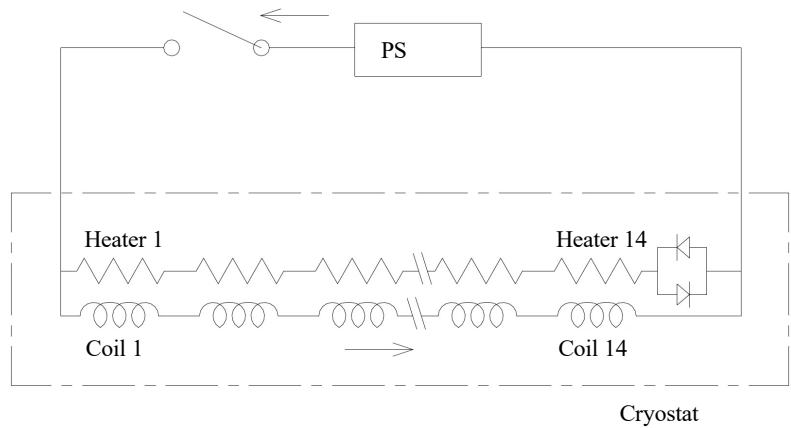

**Fig. 4.22** Quench protection circuit of COMET DS magnet.

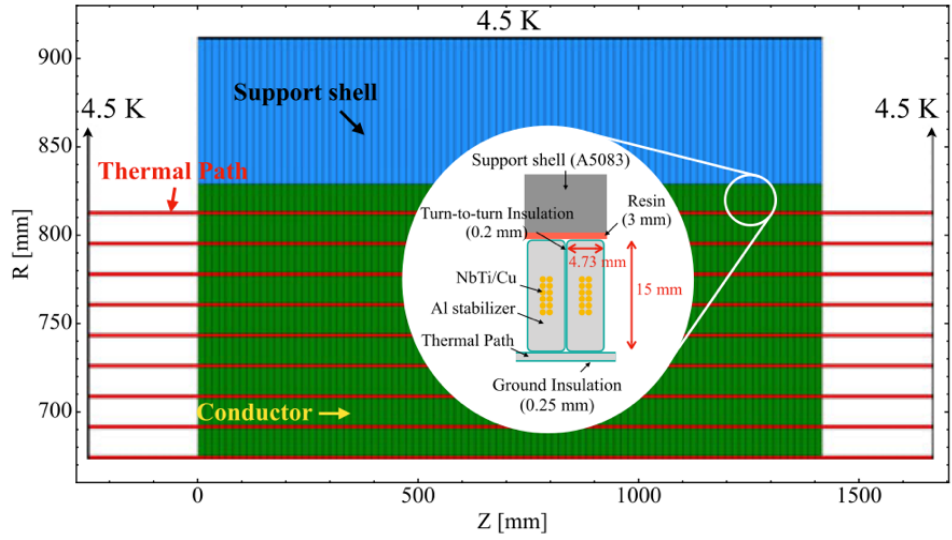

**Fig. 4.23** Cross sectional view of one PCS coil. Red, green, and blue regions indicate the aluminum strip, conductor and support shell, respectively.

**References**

[64] M. Yoshida *et al*., "Status of Superconducting Solenoid System for COMET Phase-I Experiment at J-PARC," *IEEE Trans. Appl. Supercond.*, Vol. 25, 4500904, 2015.
[65] Y. Yang *et al*., "Study of Irradiation Effects on Thermal Characteristics for COMET Pion Capture Solenoid", *IEEE Trans. On Appl. Supercond.*, Vol 28, 4001405, 2018

### 4.2.2 J-PARC g-2/EDM

In the J-PARC g-2/EDM experiment, the detector solenoid is also used as muon storage magnet. Positive muons are stored in the magnet, and decay positrons of polarized μ+ are measured. The decay positrons are detected by silicon strip detectors placed inside the muon storage orbit. One unique feature of the magnet is to adopt the three-dimensional spiral injection scheme. The muon beam enters the solenoid from the top end, and spirally go down to the storage region around the magnet center. When the beam crosses the storage region, magnetic field to kick the beam is applied to store the beam in the storage region. A very small static weak-focusing field is also applied around the storage region to maintain the beam in the storage region. Specifications required for magnetic field is summarized as follows;

- Storage regions
  - Axial magnetic field : 3T
  - Uniformity : ＜1 ppm locally, ＜0.1 ppm in circumferential average
  - Region : 33.3 ±1.5 cm in radius, ±5 cm in height
- Injection region
  - Br × Bz > 0
  - Radial field has to change smoothly along the beam orbit
  - Region : from the end of the beam injection line to the beam storage region
- Weak focus field
  - In the storage region, $B_z = B_0 - n\frac{B_{0z}}{R}r + n\frac{B_{0z}}{2R_0^2}z^2,$

- n : n-index ~ $(1.5 \pm 0.5) \times 10^{-4}$

The important feature in the superconducting magnet is the magnetic field homogeneity in the storage region, less than 1 ppm locally and 0.1 ppm in circumferential average. In order to satisfy the requirement, new analytical code using truncated singular value decomposition method to optimize coil position and size is being newly developed [66], and the design work is in progress using both the new code and existing commercial FEM code that can calculate with non-linear effect. The present design of the magnet system is shown in Fig. 4.24. The superconducting main, shim and weak focusing coils are wound with a conventional Nb-Ti wire with copper stabilizer. Main characteristics of the magnet are summarized in Table 4.4. A superconducting switch is also connected with the main coils and these are operated in persistent current mode so that the magnetic field fluctuation caused by a voltage ripple of power supply can be ignored. All superconducting coils are cooled by liquid helium in cryostat to keep coil temperature constant. GM cryocoolers are attached to the cryostat to minimize evaporation of liquid helium, and decrease the change of temperature distribution in the coils. Iron yoke covers the superconducting magnet to decrease the effect of ferromagnetic material outside the magnet on the field homogeneity, and iron yoke with cylindrical poles to make the homogeneous magnetic field in the storage region with ring shape.

The magnet is operated with persistent current mode as describe above, and it means that the stored energy must be mainly consumed during quench. In order to decrease the current decay time, and enhance the quench propagation by utilizing AC loss in superconductor, small loops are made using diodes as shown in Fig. 4.25. The loops are adjusted in such a way that the self and mutual inductance match with each other. The simulated temperature and volage in the coils are shown in Fig. 4.26. The peak temperature and voltage are calculated to be around 180 K and 1.6 kV, indicating that the magnet could be safely protected from the quench.

These 3-D magnetic field design and control technologies are being developed in collaboration with Ibaraki University. Accompanying the precise magnetic field control, precise magnetic field monitoring system development is necessary. US-JP collaboration about NMR magnetometer with ultra-high precision [67] is being progressed effectively.

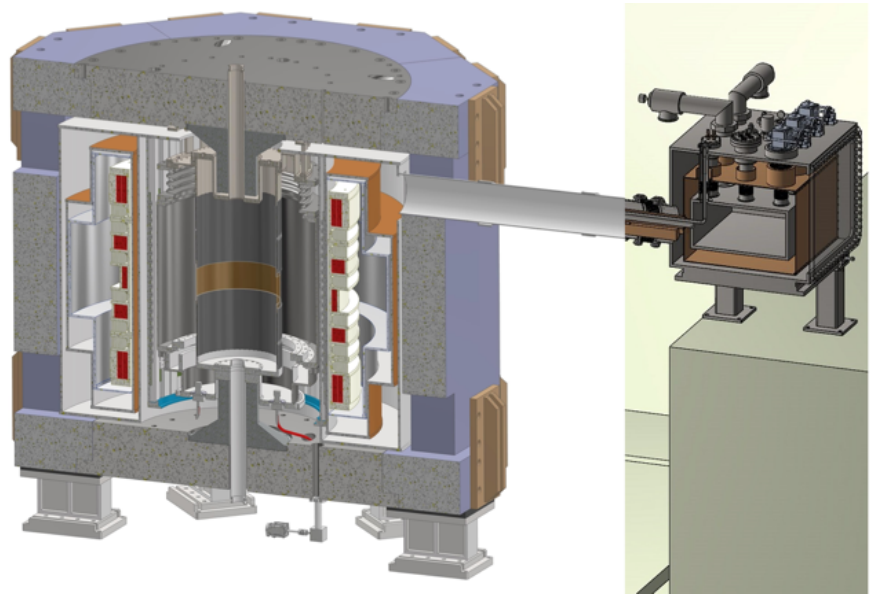

**Fig. 4.24** J-PARC g-2/EDM magnet overview.

Table 4.4 Main parameters of the storage magnet

| Item | Unit | Value |
|---|---|---|
| Nominal current | A | 423.4 |
| Stored energy | MJ | 17.2 |
| Magnet inductance | H | 198 |
| Peak field on SC coil | T | 4.9 |

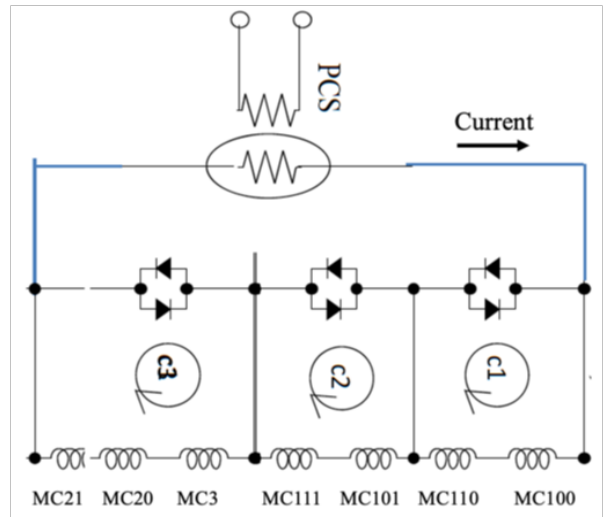

Fig. 4.25 Electrical circuit for the J-PARC g-2/EDM magnet.

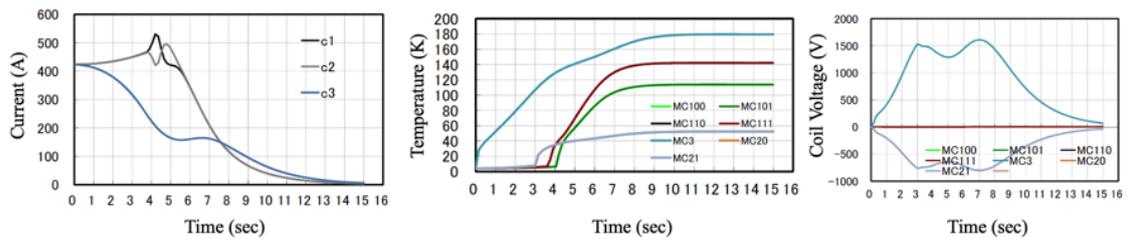

Fig. 4.26 Quench simulation results.

### References

[66] M. Abe *et al.*, "Coil Block Designs with Good Homogeneity for MRI Magnets Based on SVD Eigenmode Strength," *IEEE trans. Magn.* Vol. 51, 7002713 (2015)
[67] H. Yamaguchi *et al.*, "Development of a CW-NMR Probe for Precise Measurement of Absolute Magnetic Field," *IEEE Trans. App. Supercond.*, Vol. 29, 9000904 (2019).

## 5. Summary

Various superconducting detector solenoids for particle physics have been developed since the 1970's. A key technology is the aluminum stabilized superconducting cable for the almost all the detector magnets in particle physics experiments. With the progress of Al-stabilized conductor, the coil fabrication technology has been also progressed, that is, the inner winding technology directly inside the support cylinder, indirect cooling scheme, utilization of pure aluminum strips for the safe quench protection, and so on. The vacuum vessel design study has been also progressed, especially in transparent detector solenoids, e.g., isogrid type and honeycomb type vacuum vessels. These technologies have been used successfully for the manufacturing of many superconducting detector magnets in the past four decades thanks to the technical and scientific competencies developed, with many regular breakthroughs.

The detector solenoids design study is in progress for future big projects in Japan and Europe, that is, ILC, FCC and CLIC, based on the technologies over many years. The magnet size for each project is as large as or larger than the magnets, like CMS and ATLAS-CS, and higher strength while keeping higher RRR is a key point for the development of Al-stabilized conductor. In addition, the larger current capacity is required accompanying with the larger bore size. The ILC and FCC groups is continuing the design study of the conductor.

The present concern for the detector solenoid development is to gradually lose the key technologies and experience. Complementary efforts are needed to reach again an equivalent level of expertise, to continue the effort on research and to develop these technologies and apply them to each future detector magnet project, especially, for the development of Al-stabilized conductor fabrication. A worldwide collaboration is needed to reach and validate the required performances. KEK and CERN jointly propose to organize a workshop to share the awareness of industrial issue on Al-stabilized conductor fabrication, in which, all stakeholders will be invited, that is, superconducting magnet scientists, engineers of conductor industries and physicists who plan and design the future particle experiments.

For the detector solenoid for mid-scale experiment using a conventional copper-stabilized Nb-Ti conductor, unique features like precise control of magnetic field distribution, might be required. The development efforts are on-going in terms of the magnetic field design technology with high precision simulation, the coil fabrication technology to achieve the design requirement and the control method of magnetic field distribution.